\documentstyle[epsf,twoside]{article}

\catcode`\@=11
\def\@cite#1#2{{$^{#1}$\if@tempswa , #2\fi }}  
\def\@bite#1#2{{${#1}$\if@tempswa , #2\fi }}  
\long\def\@makefntext#1{ 
\protect\noindent \hbox to 3.2pt {\hskip-.9pt  
$^{{\eightrm\@thefnmark}}$\hfil}#1\hfill} 

\def\thefootnote{\fnsymbol{footnote}}
 \def\@makefnmark{\hbox to 4pt{$^{\@thefnmark}$\hss}}  

\def\ps@myheadings{\let\@mkboth\@gobbletwo
\def\@oddhead{\hbox{} 
\rightmark\hfil\eightrm\thepage}
\def\@oddfoot{}\def\@evenhead{\eightrm\thepage\hfil 
\leftmark\hbox{}}\def\@evenfoot{}

\def\sectionmark##1{}\def\subsectionmark##1{}}  

\oddsidemargin=\evensidemargin
\renewcommand{\thefootnote}{\fnsymbol{footnote}}

\newcounter{sectionc}\newcounter{subsectionc}\newcounter{subsubsectionc}
\renewcommand{\section}[1] {\vspace{12pt}\addtocounter{sectionc}{1} 
\setcounter{subsectionc}{0}\setcounter{subsubsectionc}{0}\noindent 
        {\tenbf\thesectionc. #1}\par\vspace{5pt}}
\renewcommand{\subsection}[1] {\vspace{12pt}\addtocounter{subsectionc}{1} 
        \setcounter{subsubsectionc}{0}\noindent 
        {\bf\thesectionc.\thesubsectionc. {\kern1pt \bfit #1}}\par\vspace{5pt}}
\renewcommand{\subsubsection}[1] {\vspace{12pt}\addtocounter{subsubsectionc}{1}
        \noindent{\tenrm\thesectionc.\thesubsectionc.\thesubsubsectionc.
        {\kern1pt \tenit #1}}\par\vspace{5pt}}

\newcommand{\textlineskip}{\baselineskip=13pt}
\newcommand{\smalllineskip}{\baselineskip=10pt}

\def\eightcirc{
\begin{picture}(0,0)
\put(4.4,1.8){\circle{6.5}}
\end{picture}}
\def\eightcopyright{\eightcirc\kern2.7pt\hbox{\eightrm c}} 

\newcommand{\copyrightheading}[1]
        {\vspace*{-1.2cm}\smalllineskip{\flushleft
        {\eightrm International Journal of Modern Physics C #1}\\
        {\eightrm $\eightcopyright$\, World Scientific Publishing
         Company}\\
         }}

\def\abstracts#1#2#3{{
        \centering{\begin{minipage}{4.5in}\baselineskip=10pt\eightrm
        \parindent=0pt #1\par 
        \parindent=15pt #2\par
        \parindent=15pt #3
        \end{minipage}
        \vspace{2pt} }\par}}

\renewenvironment{thebibliography}[1]                   
        {\ninerm 
         \baselineskip=11pt                             
         \begin{list}{\arabic{enumi}.}                  
        {\usecounter{enumi}\setlength{\parsep}{0pt}     
         \setlength{\leftmargin 17pt}{\rightmargin 0pt} 
         \setlength{\itemsep}{0pt} \settowidth          
        {\labelwidth}{#1.}\sloppy}}{\end{list}} 

\newcounter{itemlistc}
\newcounter{romanlistc}
\newcounter{alphlistc}
\newcounter{arabiclistc}

\newcounter{tempfigtabc}                        
\newcounter{temptabltabc}                       
\newcounter{tempfigltabc}                       

\def\pmb#1{\setbox0=\hbox{#1}
        \kern-.025em\copy0\kern-\wd0
        \kern.05em\copy0\kern-\wd0
        \kern-.025em\raise.0433em\box0}

\def\fnt#1#2{\footnotetext{\kern-.3em
        {$^{\mbox{\scriptsize #1}}$}{#2}}}

\def\runninghead#1#2{\pagestyle{myheadings}
\markboth{{\eightit{\quad #1}}\hfill}{\hfill{\eightit{#2\quad}}}}
   
\font\tenbf=cmbx10
\font\tenit=cmti10 
\font\tenit=cmti10
\font\bfit=cmbxti10 at 10pt
 1
 1
 1

\font\ninerm=cmr9

\font\eightrm=cmr8
\font\eightit=cmti8

\renewcommand{\thefootnote}{\fnsymbol{footnote}} 
\def\bsc{{\sc a\kern-6.4pt\sc a\kern-6.4pt\sc a}}  
\def\bflatex{\bf L\kern-.30em\raise.3ex\hbox{\bsc}\kern-.14em 
T\kern-.1667em\lower.7ex\hbox{E}\kern-.125em X}

\def\@citex[#1]#2{%
\if@filesw \immediate \write \@auxout {\string \citation {#2}}\fi
\@tempcntb\m@ne \let\@h@ld\relax \def\@citea{}%
\@cite{%
  \@for \@citeb:=#2\do {%
    \@ifundefined {b@\@citeb}%
      {\@h@ld\@citea\@tempcntb\m@ne{\bf ?}%
      \@warning {Citation `\@citeb ' on page \thepage \space undefined}}%
      {\@tempcnta\@tempcntb\advance\@tempcnta\@ne%
      \@tempcntb\number\csname b@\@citeb\endcsname\relax%
      \ifnum\@tempcnta=\@tempcntb 
        \ifx\@h@ld\relax
          \edef \@h@ld{\@citea\csname b@\@citeb\endcsname}%
        \else
          \edef\@h@ld{\ifmmode{-}\else--\fi\csname b@\@citeb\endcsname}%
        \fi
      \else   
        \@h@ld\@citea\csname b@\@citeb \endcsname
        \let\@h@ld\relax
      \fi}%
    \def\@citea{,\penalty\@highpenalty\,}%
  }\@h@ld
}{#1}}

\DeclareRobustCommand\bite{%
  \@ifnextchar [{\@tempswatrue\@bitex}{\@tempswafalse\@bitex[]}}
\def\@bitex[#1]#2{%
  \let\@bitea\@empty
  \@bite{\@for\@biteb:=#2\do
    {\@bitea\def\@bitea{,\penalty\@m\ }%
     \edef\@biteb{\expandafter\@firstofone\@biteb}%
     \if@filesw\immediate\write\@auxout{\string\citation{\@biteb}}\fi
     \@ifundefined{b@\@biteb}{\mbox{\reset@font\bfseries ?}%
       \G@refundefinedtrue
       \@latex@warning
         {Citation `\@biteb' on page \thepage \space undefined}}%
       {\hbox{\csname b@\@biteb\endcsname}}}}{#1}}

\newcommand{\beq}{\begin{equation}}
\newcommand{\eeq}{\end{equation}}        
\newcommand{\bqa}{\begin{eqnarray}}        
\newcommand{\eqa}{\end{eqnarray}}        
\newcommand{\be}{\begin{enumerate}}
\newcommand{\ee}{\end{enumerate}}        
\newcommand{\bi}{\begin{itemize}}
\newcommand{\ei}{\end{itemize}}        
\newcommand{\fig}[1]{Fig.~\ref{#1}}
\newcommand{\alg}[1]{Algorithm~\ref{#1}}

\renewcommand{\matrix}[1]{\mbox{\sf #1}}

\newcommand{\eg}{{\frenchspacing\em e.\hspace{0.4mm}g.{}}}
\newcommand{\ie}{{\frenchspacing\em i.\hspace{0.4mm}e.{}}}
\newcommand{\eq}[1]{{\frenchspacing Eq.~\ref{#1}}}
\newenvironment{algo}[3]{\vglue6pt\noindent
  \framebox{\begin{minipage}{.97\textwidth}\begin{algorithm} #2 \rm #1\par
        \footnotesize\begin{center}\begin{minipage}{.97\textwidth}
\begin{tabbing}  #3
\end{tabbing}\end{minipage}\end{center}\end{algorithm}\end{minipage}}\vglue6pt}{}
\newtheorem{algorithm}{Algorithm}

\runninghead{Th.\ Lippert et al.} 
{FFT for the APE Parallel Computer}

\begin{document}
\textlineskip
\setcounter{page}{1}
\thispagestyle{empty}

\copyrightheading{}

\vspace*{0.88truein}
\centerline{\bf FFT for the APE Parallel Computer}
\vglue 28pt
\centerline{\footnotesize 
$^a$THOMAS LIPPERT, $^a$KLAUS SCHILLING, $^{b,c}$FEDERICO TOSCHI,} 
\vglue8pt
\centerline{\footnotesize 
$^a$SVEN TRENTMANN, $^c$RAFFAELE TRIPICCIONE}
\vglue 12pt
\centerline{\footnotesize\it $^a$HLRZ c/o KFA-J\"ulich and DESY}
\baselineskip=10pt
\centerline{\footnotesize\it D-52425 J\"ulich, Germany} 
\vglue 12pt
\centerline{\footnotesize\it $^b$Dipartimento di Fisica, Universit\`{a} di Pisa} 
\baselineskip=10pt
\centerline{\footnotesize\it Piazza Torricelli 2, I-56126, Pisa, Italy,} 
\centerline{\footnotesize\it and}
\centerline{\footnotesize\it INFM, Sezione di Tor Vergata,}
\baselineskip=10pt
\centerline{\footnotesize\it Via della Ricerca Scientifica 1, I-00133, Roma, Italy.} 
\vglue 12pt
\centerline{\footnotesize\it $^c$ INFN, Sezione di  Pisa, c/o} 
\baselineskip=10pt
\centerline{\footnotesize\it I-56010 S. Piero, Pisa, Italy} 
\vglue 17pt
\centerline{\eightrm Received }
\vglue 17pt

\setcounter{footnote}{0}
\renewcommand{\thefootnote}{\alph{footnote}}
\abstracts{\noindent We present a parallel FFT algorithm for SIMD systems
  following the `Transpose Algorithm' approach.  The method is based on the
  assignment of the data field onto a 1-dimensi\-onal ring of systolic
  cells.  The systolic array can be universally mapped onto any parallel
  system.  In particular for systems with next-neighbour connectivity our
  method has the potential to improve the efficiency of matrix
  transposition by use of hyper-systolic communication.  We have realized a
  scalable parallel FFT on the APE100/\-Qua\-drics massively parallel
  computer, where our implementation is part of a 2-dimensional
  hydrodynamics code for turbulence studies.}{}{\vglue8pt\noindent
  }

\section{Introduction}
\normalsize
\textlineskip
\noindent
The efficient implementation of Fast Fourier Transforms (FFT) on
parallel high performance computers is of tantamount importance in
order to warrant broad scientific and industrial application of these
systems.

The ongoing development of new high speed low cost parallel computers
with Teraflops performance, designed primarily to solve problems in
lattice quantum field theory\cite{APE100,APE1000,CHRIST1,CHRIST2},
generates interest in other scientific and commercial fields with
Tera-computing needs. There are quite a few such applications for
which one would like to spend 10 to 1000 sustained Teraflopshours.
However, low cost parallel systems go along with certain architectural
restrictions as \eg\ next-neighbour connectivity\cite{SIMD,GOTTLIEB}.
Numerical procedures like FFT that involve non-local inhomogeneous
interprocessor communication patterns are therefore hard to implement
efficiently on these machines.  The communication bottleneck of FFT is
even more serious on single addressing machines like APE100, referred
to as SISAMD in the following\footnote{SISAMD: \underline{S}ingle
  \underline{I}nstruction \underline{S}ingle \underline{A}dressing
  \underline{M}ultiple \underline{D}ata}, where the individual
processing units do not provide indexed local addressing\cite{APE100}.

The substantial cost-factor of FFT on parallel computers arises from
`bit-inversion' of the location of each input data element and from
the pair-wise decimation of the bit-inverse ordered array
elements\cite{NUMRES}.  As a result, FFT is harassed by frequent
non-contiguous memory accesses. On a standard computer, this fact
implies repeated cache misses, while on a parallel machine
it translates into non-local interprocessor communication.

A well known strategy to carry out FFT on parallel systems for a
dimensionality $\ge 2$ is based on the splitting of the
$d$-dimensional FFT into $d$ 1-dimensional FFT's\cite{NUMRES}.  This
approach\footnote{A second standard procedure, the `Bit-Exchange'
  algorithm, is better suited for hyper-cubic architectures than mesh
  computers.} is known as the `Transpose Algorithm'\cite{KUMAR}.  Let
us start from a $d$-dimensional data field with coordinates
$i_1,i_2,\dots,i_d$, $i_k=0,\dots,n_{k}-1$.  The field may be
distributed such that one axis (coordinate $i_1$) is spread across the
processors, while the other axes are locally assigned to the
processors.  In this case, the 1-dimensional FFT can be computed on
the local axes (coordinates $i_2,\dots,i_d$), \ie\ for each coordinate
$i_k$ in parallel without access to remote memory.  However, as at
least one axis has to be distributed across the processors, a
reordering of the data field between distributed and local axes is
required. Such a reordering of the data elements is
equivalent to a matrix transposition. So the generic task is to devise
an efficient Transpose Algorithm for distributed array
layout\cite{KUMAR}.

In Ref.~\bite{CABIBBO}, a promising implementation of the Transpose
Algorithm, adapted to mesh-based SIMD computers with next-neighbour
connectivity, has been discussed. A $k$ dimensional field of volume
$V$ is assigned to the processors within a $d$-dimensional toroidal
mesh network.  The transposition of the field is performed in a
three-step procedure: first the data field is skewed (like rendering a
matrix skew).  This step is followed by interprocessor communication.
The final step is an inverse skewing.  

The authors of Ref.~\bite{CABIBBO} specify a pattern for the interprocessor communication
adapted to $d$-dimensional mesh architectures like APE.  For the
3-dimensional connectivity of APE100/Quadrics they reached the
gratifying result that the interprocessor communication can be
performed with complexity of $V/p^{\frac{2}{3}}$ data communication
operations per processor, where $V$ is the data volume.

The two local skewing steps at the beginning and the end of the
transposition have to be carried out according to the machine's
addressing capabilities: on SIMD or MIMD machines, skewing is nothing
but a local data assignment.  An extra difficulty arises on SISAMD
computers since they lack local addressing. The lack of local
addressing at first sight renders the skewing phase {\em
  non-scalable}, \ie, each processor has to carry out its copy
operations separately.  This leads to an unacceptable performance
degradation on SISAMD machines\footnote{ Ref.~\bite{CABIBBO}
  discusses the implementation of the Transpose Algorithm for
  Multi-controller APE systems that can partly overcome the SISAMD
  restrictions.}. We will largely overcome this problem, 
by introducing a novel skewing algorithm that proceeds in $\log p$ steps.

Starting out from Ref.~\bite{CABIBBO}, we construct a general transpose
algorithm for FFT, that can be implemented in a scalable manner on any
parallel architecture.  For this purpose, we first associate the data
field to a 1-dimensional ring of systolic cells\cite{PETKOV}, that can
perform both computations and communication (to neighbouring cells).
Our systolic ring can be readily mapped onto any parallel machine's
processor connectivity.  One way to do this for APE100/Quadrics has
been presented in Ref.~\bite{HYPER_IJMPC}.  We can reduce the time
expense of systolic interprocessor communication along the systolic
ring by application of hyper-systolic data movement\cite{HYPER,PDPTA}.
Implemented on a $d$-dimensional mesh, each processor has to perform
$O(V/p^{\frac{d-1}{d}})$ communication operations on average.

One particular target implementation system of our FFT  will be the
European Teraflops system of 1998, APEmille\cite{APE1000}.  APEmille is
designed as SIMD machine with next-neigh\-bor connectivity.

There are of course many Tera-computing applications where efficient
parallel FFT algorithms will be of use, \eg\ spectral methods in global
climate modeling, turbulent fluid dynamics or quantum chromodynamics.  The
latter two will serve us as examples for illustration and benchmarking of
the Transpose Algorithm.

The implementation of a 1-dimensional FFT on APE which is important
for signal processing applications will be presented in an upcoming
publication\cite{ONED}.

\section{Transpose Algorithm on a Systolic Ring}
\noindent
We construct a Transpose Algorithm as implemented on a 1-dimensional
systolic array with toroidal topology.

\subsection{Data assignment}
\noindent
Let us consider a 2-dimensional data field (matrix) $\matrix{M}$ of size
$p\times p$, where $p$ is the number of processors of the parallel system.
Each element $m_{ji}$, $i=0,\dots,p-1$, $j=0,\dots,p-1$, can be considered
as sub-matrix. This structure will allow for easily generalization to data
arrays of size $V=m\times n=\frac{m}{p}\times\frac{n}{p} \otimes p\times p$
with rectangular shape.

In the following, we assume that the implementation network supports the
embedding of a 1-dimensional ring topology. Along the ring, circular shifts
in an homogeneous pattern can be performed.

\begin{figure}[!htb]
\centerline{\epsfxsize=.45\textwidth\epsfbox{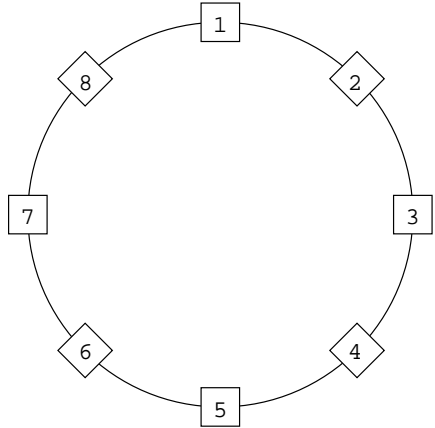}}
\caption[Layout]{
Systolic array of 8 systolic cells forming a systolic ring.
\label{SYSTOLICRING}}
\end{figure}

To achieve a universal Transpose Algorithm we are going to discuss the
method as implemented on an array of systolic cells that can perform
compute, store and communication operations.  For illustration, in
\fig{SYSTOLICRING}, a $1$-dimensional ring with $p=8$ cells is sketched.
The cells are connected to their next neighbours by direct communication
lines.
The cells can perform communication operations to the next neighbours in a
homogeneous pattern as well as synchronous compute operations with equal
load, on cell resident data.  This systolic array concept is simple enough
to be mapped efficiently onto practically any parallel architecture and thus is the
adequate framework for the construction of a ``universal'' Transpose Algorithm.

\begin{figure}[!htb]
\centerline{\epsfxsize=.8\textwidth\epsfbox{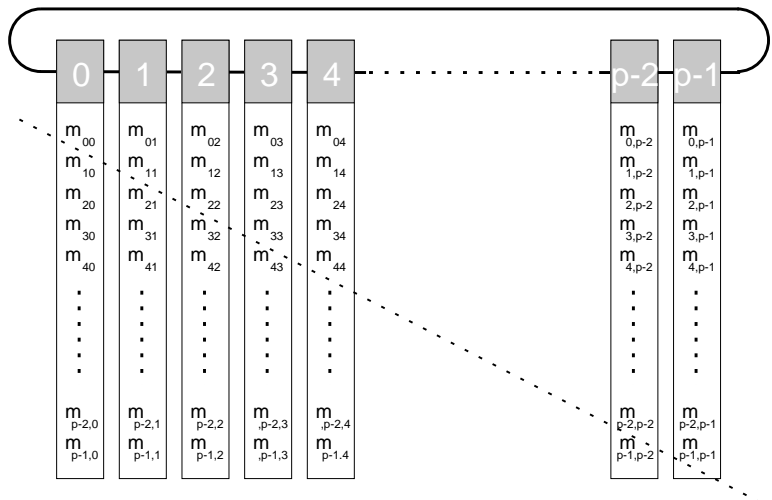}}
\caption[Layout]{
  Assignment of  the $p\times p$ data array $\matrix{M}$
  to the $p$-cell ring.
\label{LAYOUT_NORMAL}}
\end{figure}

The initial data assignment on the systolic ring is given in
\fig{LAYOUT_NORMAL}. The data elements are indexed following standard
matrix notation.  Note that 
matrix elements with equal second coordinate $i$
are all assigned to  cell $i$, while their first coordinate refers to 
their local address.

\subsection{Transposition}
\begin{figure}[tb]
\centerline{\epsfxsize=.8\textwidth\epsfbox{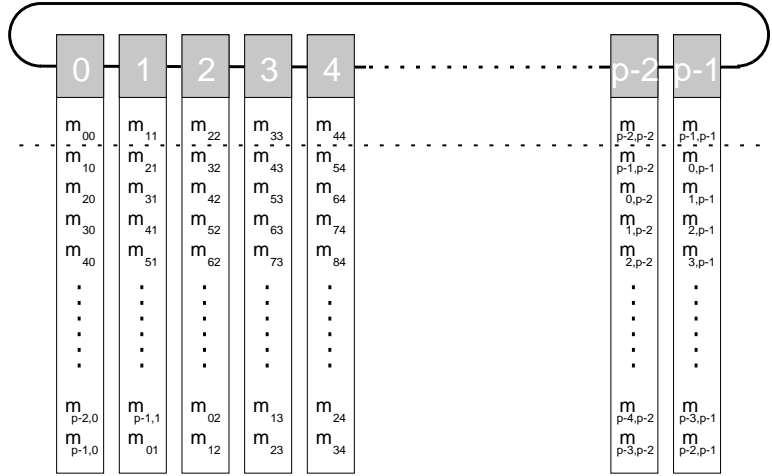}}
\caption[Layout]{
  Distribution of a 2-dimensional $p\times p$ data array $\matrix{M}$
  on a $p$-cell ring in skew order.
\label{LAYOUT_SKEW}}
\end{figure}

\noindent
The transposition of $\matrix{M}$ proceeds within 3 steps, a pre-skewing
step, that does not involve interprocessor communication, a row-shifting
step, requiring interprocessor communication and finally a re-skewing
step, again without interprocessor communication:
\paragraph{1. Pre-Skewing.} 
The matrix $\matrix{M}$ is locally skewed along the columns.  The
$i$-th column is shifted upward by a circular shift of stride $i$.
After finishing this process for all columns $i$, the diagonal
elements of the original matrix are located in the first row, see
\fig{LAYOUT_SKEW}.

The skewing procedure is given in the pseudo-code
representation of \alg{SKEWING}.
\begin{algo}{Pre-Skewing.}%
            {\label{SKEWING}}{ 
for\= for\= for\= for\= for\= for\=\kill 
foreach cell $i=0:p-1$ $\in$ systolic array \\
\> for $j=0:p-1$ \\ 
\> \> $m_{j,i}=m_{[(j+i)\mbox{\footnotesize mod}_{p}],i}$ \\ 
\> end for \\ 
end foreach}
\end{algo} 
Note, that in each step different cells act on different internal
addresses. As all cells of a systolic array perform equal operations per
definition, skewing is not a systolic process! However, we will demonstrate
in appendix~A how skewing can be performed in form of a
systolic process, within $p\log{p}$ steps. Therefore, by mapping of the
systolic array onto a SISAMD systems we are able to perform skewing in a
scalable manner.  Of course on SIMD and MIMD systems skewing is trivial as
it can be carried out by usage of local addressing facilities.

\paragraph{2. Row-shifting.}  
In the second phase, the rows of $\matrix{M}$ are shifted along the
ring.  After the shift, the rows are assigned to a  local memory
location determined as follows: the $j$-th row is circularly shifted by
a distance of $j$ cells to the right and subsequently is copied from
storage location $j$ to storage location $(p-j){\mbox{\footnotesize
    mod}}_{p}$.  The procedure is given in \alg{HYPERCOM}.

\begin{algo}{Row-Shifting.}%
            {\label{HYPERCOM}}{
for\= for\= for\= for\= for\= for\=\kill 
foreach cell $i=0:p-1$ $\in$ systolic array \\
\> for $j=0:p-1$ \\ 
\> \> $m_{j,i}=m_{
[(p-j)
\mbox{\footnotesize mod}_{p},
(i+j)
\mbox{\footnotesize mod}_{p}]
}$ \\ 
\> end for \\ 
end foreach}
\end{algo} 

\begin{figure}[!htb]
\centerline{\epsfxsize=.8\textwidth\epsfbox{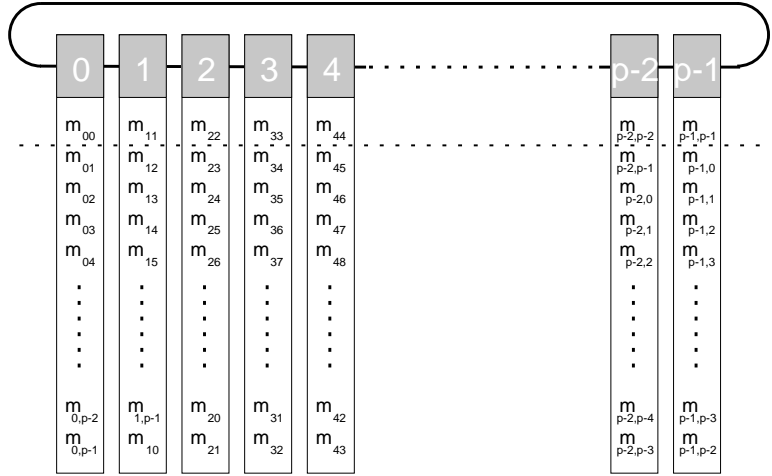}}
\caption[Layout]{
  $p\times p$ data array $\matrix{M}$ after row-shifting.
  \label{LAYOUT_HYPER}}
\end{figure}

The corresponding data locations after completion of phase 2 are
illustrated in \fig{LAYOUT_HYPER}.

We emphasize that in any step, each cell acts on the same local addresses.
Therefore, the row-shifting is a systolic process and, mapping the systolic
array onto the processors of a parallel computer, local address
computations are not required.

\begin{figure}[!htb]
\centerline{\epsfxsize=.8\textwidth\epsfbox{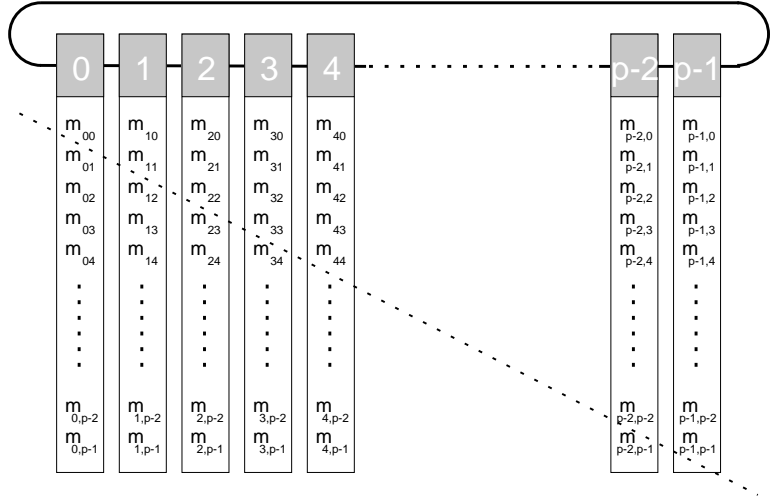}}
\caption[Layout]{
  Transposed 2-dimensional $p\times p$ data array $\matrix{M}$ on a
  $p$-cell ring.
\label{LAYOUT_TRANS}}
\end{figure}

\paragraph{3. Re-Skewing.} 
In a last step, the matrix $\matrix{M}$ is skewed downwards, in analogy to
the pre-skewing step.  The $i$-th column is shifted by a circular shift of
stride $-i$.  The final data arrangement is depicted in \fig{LAYOUT_TRANS}.

The pseudo-code representation of the re-skew step is given in
\alg{RESKEWING}.
\begin{algo}{Re-Skewing.}%
            {\label{RESKEWING}}{
for\= for\= for\= for\= for\= for\=\kill 
foreach cell $i=0:p-1$ $\in$ systolic array \\
\> for $j=0:p-1$ \\ 
\> \> $m_{j,i}=m_{[(j-i+p)\mbox{\footnotesize mod}_{p},i]}$ \\ 
\> end for \\ 
end foreach}
\end{algo} 

For  FFT in more than 1 dimensions, two such  transposition have
to be carried out. In \alg{FFT}, we give a 
pseudo-code template for parallel FFT in two dimensions:
\begin{algo}{2-dimensional parallel FFT.}%
            {\label{FFT}}{
for\= for\= for\= for\= for\= for\= for\=\kill 
call 1-d-FFT \>\>\>\>\> ! on local axis\\ 
call transpose\\
\> call pre-skew\\
\> call row-shift\\ 
\> call re-skew\\
call 1-d-FFT \>\>\>\>\> ! on local axis\\ 
call transpose\\
\> call pre-skew\\
\> call row-shift\\
\> call re-skew}
\end{algo} 
For the description of FFT we refer to Ref.~\bite{NUMRES}.

\section{Mapping on APE100/Quadrics}
\noindent
The ring of systolic cells together with their functionality can
readily be mapped onto the processors of a parallel machine that
supports a 1-dimensional ring topology. We are going to describe the
mapping for APE100/Quadrics.

\subsection{The APE100/Quadrics parallel system} 
\noindent
The parallel system APE100 has been designed primarily for quantum
chromody-nam\-ics\cite{ROTHE} computations on a lattice. Its commercial
version goes under the name Quadrics.  The diagram \fig{QUAD} depicts
the Quadrics system:
\begin{figure}[!htb]
\centerline{\epsfxsize=.9\textwidth\epsfbox{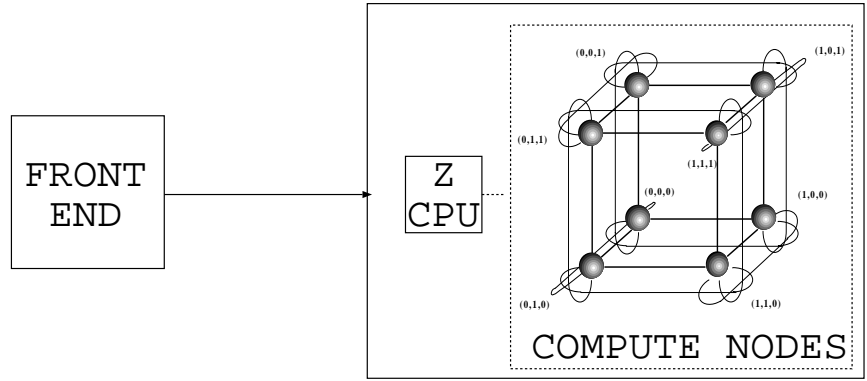}}
\caption[Quadrics]{
The Quadrics system.
\label{QUAD}}
\end{figure}
compilation and program submission as well as I/O tasks are
carried out on a front-end workstation.  The so-called Z-CPU
manages the instruction stream (synchronous for all computing
nodes) and performs global integer arithmetics.

{The programming language TAO} looks FORTRAN like. It supports so-called
syntagmas, structs which are very useful for efficient programming.  TAO 
supports a local address space whereas communication from and
to remote memories uses additional global address bits.

{The compute nodes} perform the local floating point operations. They are
composed of memory, a floating point unit called MAD, and the communication
chips.  The MAD is able to perform 2 floating point operations within one
clock cycle, that is $(25\, \mbox{MHz})^{-1}$.  Thus, one processor has a peak speed of 50 Mflops.

The smallest configuration of the Quadrics computers is a Q1, a unit
of 8 compute nodes, called a board.  The nodes are connected by 6
communication links to their neighbor nodes in a 3-dimensional grid,
see \fig{QUAD}. Larger machines like the 32-node Q4 or the 512-node
QH4 are composed of several boards. The network topology of the large
system is hard-wired. A 512-node Quadrics QH4 is typically configured
as an $8\times 8\times 8$-grid, toroid\-ally closed in each direction.

In order to transfer one 32-bit word from the memory of the
neighboring processor to the memory of the local processor, apart from
start-up times, 8 clock cycles are needed in the ideal case. In
practical applications, this limit can not be reached completely,
however.

APE100/Quadrics is the European pathway towards Tera-computing. The APE
groups in Pisa and Roma are currently developing their next generation of
SIMD machines, the parallel APEmille systems. This machine will exhibit a compute
performance of more than 1 Teraflops.  The commercial version of this
machine will go under the name PQE2000.  This machine will have a parallel
MIMD front-end system.

The APEmille architecture is extremely similar to APE100 (same topology and interconnectivity). Improvements, beyond much higher performances, include integer and double precision arithmetic.


As the most important extension to the present APE100 architecture all
processing nodes may access their own local memories using different
local addresses.  Thus, APEmille will overcome the local addressing
restrictions of APE100, allowing to implement algorithms that could not
scale on APE100 machines.  

A second advantage is the more general long-distance routing
capability.  Indeed, rigid communications can be set-up at arbitrary
distances and automatically handled by the network
(cut-through-routing). Again, this simplifies coding of the systolic
moves and reduces communication latency.  Furthermore, the APE100
software will be upward compatible for APEmille, and FFT-packs
developed for APE100 can readily be ported onto APEmille.

\subsection{Embedding a 1-dimensional ring on Quadrics}
\noindent
As has been shown in detail in Refs.~\bite{HYPER_IJMPC} and
\bite{CETRARO}, the 3-dimensional network of the
Quadrics can be exploited to emulate the 1-dimensional ring
functionality and to carry out shifts of any stride. 

On a 3-dimensional grid based machine, the shifts can be described by a
sequence of hardware communication operations in each of the three
directions.  We encode the stride $a_{l}$ in terms of the distance between
processor $0$ and processor $l$ on the systolic ring, by
\begin{equation}
l=l_{2} + l_{1} p_{2}+l_{0} p_{2}p_{1}=a_l,
\end{equation}
with $p_i$ being the number of processors in direction $i$, and $l_i$ the
coordinates of processor $l$ in direction $i$ on the 3-dimensional
processor grid, the total number of processors being $p=p_{0}p_{1}p_{2}$.
The cost function $C(a_l)$ in units of hardware communication operations is
given as
\begin{equation}
C(a_{l})=l_{0}+l_{1}+l_{2}.
\end{equation}
The total cost function $C_t$ is the sum of the individual $C(a_{l})$
over all circular shifts with stride $a_l$, $a_l=0,\dots,p-1$:
\begin{equation}
C_t=
\sum_{l_2=0}^{p_2-1}
\sum_{l_1=0}^{p_1-1}
\sum_{l_0=0}^{p_0-1}
(l_2+1+l_1+1+l_0+1) =
\frac{p}{2}(p_{0}+p_{1}+p_{2}+3).
\end{equation}

In practise, one can exploit hardware communication for each axis in both
directions.  Hence, the average cost per shift can be reduced by a factor
of 2, and on a 3-dimensional machine with next-neighbour connectivity the
wall clock time for interprocessor communication decreases like
\begin{equation}
t_{ipc}\propto\frac{1}{p}(p_0+p_1+p_2+6),
\label{SCALING}
\end{equation}
for a fixed size $V$ of the data array.

\section{Benchmarking the Transpose  Algorithm on APE100/Quadrics}
\noindent
We have mapped the FFT based on the systolic Transpose Algorithm onto
several Quadrics parallel SISAMD computers. Here we present timing results
and scaling. In the next section we discuss two physics case studies where
FFT can play an important r\^ole.

\begin{figure}[!htb]
  \centerline{\epsfxsize=.9\textwidth\epsfbox{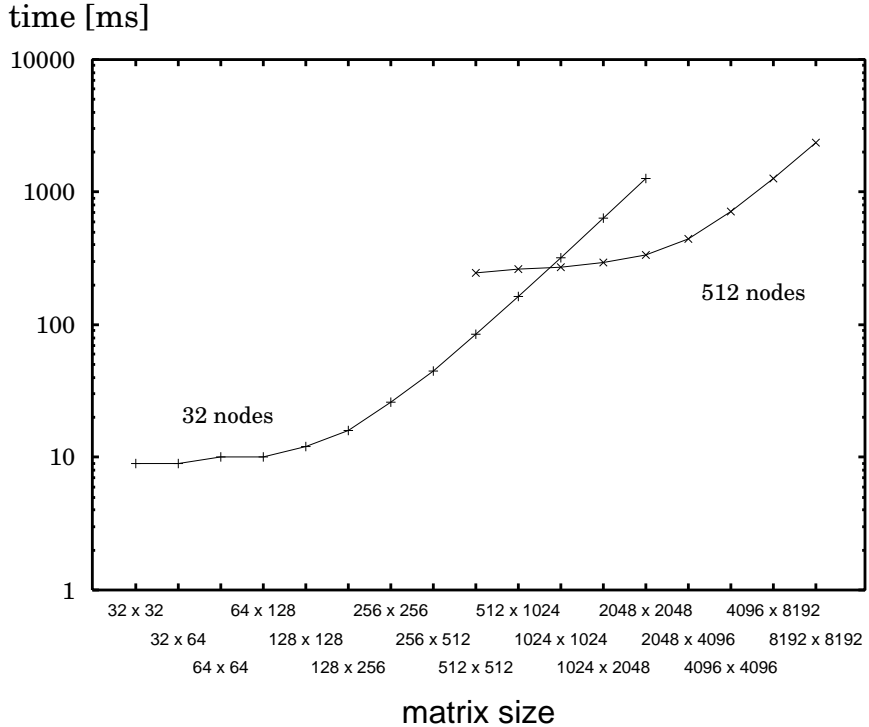}}
\caption[Layout]{Double $\log$ plot of 
transposition vs.\ matrix size $V$ on a 32-node Quadrics Q4
  and a 512-node Quadrics QH4.
\label{SIZE}}
\end{figure}

For a 32-node APE100/Quadrics Q4 system and a 512-node QH4 system we give
detailed timing results.  We plot the total execution times against the
matrix size $V$, {\em cf.\ } \fig{SIZE}.  On both machines the asymptotic
behaviour is reached when the length of the stream set involved in row-shifting is larger than 32 data elements.
For stream smaller than 32 elements we are faced with a start-up time
effect.  The two curves intersect each other for about
$n=1024$.

\begin{figure}[!htb]
  \centerline{\epsfxsize=.9\textwidth\epsfbox{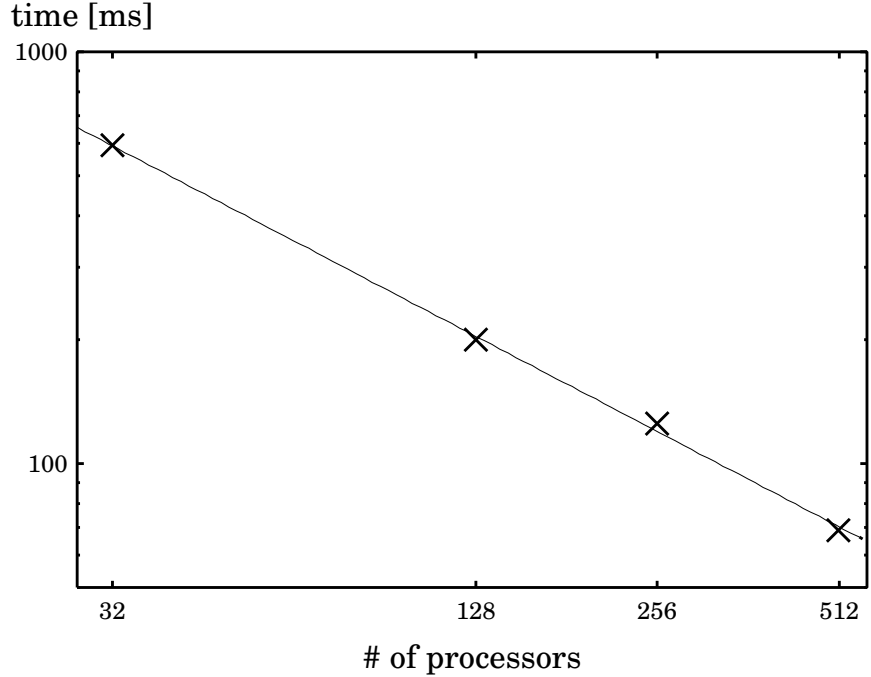}}
\caption[Layout]{
  Scaling of the Transpose Algorithm on 32-node to 512-node Quadrics
  systems for fixed problem size ($V=2048\times 2048$ elements).
\label{LAW}}
\end{figure}

In order to test the scaling law, \eq{SCALING}, we plot the times for
interprocessor communication for a fixed matrix size of $2048\times 2048$
complex elements on various Quadrics platforms, against the number of
processors (see \fig{LAW}).  A fit to the data according to $t\propto
p^{\alpha}$ reveals an exponent $\alpha$ of $-0.76$. The naive expectation
from \eq{SCALING} for cubic machine topologies would be $-0.66$.
However, the actual topologies were $V_{(p=512)}=8\times 8\times 8$,
$V_{(p=256)}=8\times 8\times 4$, $V_{(p=128)}=8\times 4\times 4$, and
$V_{(p=32)} =8\times 2\times 2$.

\section{Applications of FFT on APE100/Quadrics}
\noindent
The work described in the previous section has been motivated by the need to have a fast FFT in some physics applications
of ours.  The solution of these problems can benefit considerably from the
power of APE computers.

\subsection{2-dimensional FFT: Kraichnan model}
\noindent
In order to simulate the Kraichnan model\cite{KRAICHNAN}, an efficient
2-dimensional FFT transform was required.  The Kraichnan model is a toy
model which has attracted much interest recently, because of its analytical
simplicity with respect to the full Navier-Stokes equations.  An often
studied problem of turbulence is the anomaly of the scaling exponents in
the inertial range. In order to get reliable numerical values for these
exponents, huge computer simulations are required.

This toy model describes a passive scalar $T({\bf x},t)$ which moves
in a velocity field which is synthetically generated with a given
scaling exponent $\xi$. One studies the scaling exponent $\xi_T$ of
the passive scalar as a function of $\xi$.  The relevant equations
are:
$$\partial_t T + \left( {\bf u} \cdot \nabla \right) T =
\chi \nabla^2 T + f$$
$$\left\langle f({\bf x},t) f({\bf y},t')\right\rangle = \delta (t-t') C(\left| {\bf x}-{\bf y}\right|)$$
\begin{equation}
\label{vel_eq}
\left\langle v_i({\bf x},t) v_j({\bf y},t') \right\rangle = D^{\alpha \beta}({\bf x-y}) \delta (t-t').
\end{equation}

$$D^{\alpha \beta}(x)=D(0) \delta^{\alpha \beta} -d^{\alpha \beta}(x)$$
$$d^{\alpha \beta}(x) = D \left( \left(d+ \xi -1 \right) \delta^{\alpha \beta} - \xi {{x^{\alpha} x^{\beta}} \over {\left| x \right|^2}} \right) \left| x \right|^{\xi} $$
 
The random force $f({\bf x},t)$ is delta correlated in time and Gaussian
with correlator $C$ (a function with support of the order of size of the
system) in space. The velocity field is ${\bf u}$, its statistical
properties are given by \eq{vel_eq} and, as it is clear from the expression
for $D^{\alpha \beta}$, the incompressibility condition is fulfilled. The
exponent $\xi$ is just the scaling exponent imposed on the velocity field
(relevant values are in the range $[0,2]$).

The main practical problem for the simulation of a system of this kind are
the huge size of the system (\ie\ $8192\times 8192$) and the fact that a new velocity field is
required at each time step.

The velocity field with the required statistical properties given by
\eq{vel_eq} is easily generated in Fourier space just by extracting random
gaussian amplitudes (for the k modes) with given variance $\sim
k^{-1-\xi/2}$. The direction of the {\it k}-th component ${\bf u}_k$ is
choosen such that ${\bf k} \cdot {\bf u}_k=0$.  In addition, with FFT,
periodic boundary conditions are automatically fulfilled.

Alternative algorithms could be used for the generation of the velocity
field, but, since the required statistical property is non-local, they
would face the non-locality problem anyway and communication between
different processors would become very heavy.  In the case of our problem,
a further major performance gain can be achieved in a very simple way.
Since we need a specific geometric mapping of the variables only in real
space, we can skip a transposition step by generating transposed amplitudes
in {\it k} space. This is of course true in several interesting cases, when
we need to go to {\it k} space in order to make some computation and then
we want to come back to real space.

The FFT which has been described in this article has been implemented on a
512-node QH4 machine and has been optimized for working with very large
lattice sizes (\ie\ $8192 \times 8192$ elements).  We are able to compute
 an FFT on the corresponding date set of about 512 Mbyte in 4 seconds.  This means that we are
working at about $ 9 \%$ of the peak performance for the
above problem.  If we perform the complete FFT including the
back-transposition, we achieve about 6 \% of the peak performance.
\newpage
\subsection{4-dimensional FFT: Quantum chromodynamics}
\noindent
In principle, the above Transpose approah to parallel FFT could be applied
for systems of any dimensionality $\ge 2$. However, from a practical point
of view, in the implementation of FFT on parallel systems we meet rather
different situations: on a low dimensional lattice, the length of the axes
obviously can be much larger than that of a high dimensional lattice.
However, the length of at least two axes must be as large as the length of
the systolic ring, otherwise we would have to pad with zeroes and would
loose efficiency.

This situation occurs \eg\ in simulations of lattice quantum
chromodynamics\cite{ROTHE} (QCD). QCD is formulated on a 4-dimensional
euclidean space-time lattice.  Some aspects of QCD simulations, like
\eg\ gauge fixing\cite{QCD}, Fourier acceleration of Krylov
solvers\cite{DAVIES} or accelerated updates might involve
FFT\cite{UPDATE}.

We can not go into detail of lattice QCD here, but rather discuss the
FFT implications of a 4-dimensional field
$f(x_{0},x_{1},x_{2},x_{3})$.  This field is assigned onto a finite
4-dimensional lattice of size $x_{0}=0,\dots p_{0}-1$, $x_{1}=0,\dots
p_{1}-1$, $x_{2}=0,\dots p_{2}-1$, and $x_{3}=0,\dots p_{3}-1$.
Typical sizes of QCD lattices range between $16^{4}$ and
$64^{4}$ sites, hence on massively parallel systems with $p>100$, the
direct application of the approach described above would not be adequate for these
applications.

However, the systolic FFT idea can be readily generalized: In the case
of the 2-dimensional FFT on a 3-dimensional grid computer, one axis
was distributed while the other was local on the systolic ring.
Handling the 4-dimensional FFT on a 3-dimensional grid computer one
has more than one option.

A very natural choice is to stick to the lexicographic data layout of
the 4-dimensional QCD lattice.  This leads to introduce {\em many}
systolic rings along the grid axes of a given dimension.  On a
3-dimensional grid the set of axes along a given direction is
specified by fixing two grid coordinates orthogonal to the direction,
say coordinates $x_{0}$ and $x_{1}$.  Then, we can consider the axes
specified by $x_{0}$ and $x_{1}$ of our 4-dimensional system as a
number of $p_{0}p_{1}$ systolic rings. On each ring, coordinate
$x_{2}$ of the 4-d system is laid out while $x_{3}$ is local.
Therefore, each of the rings corresponds to the situation we have
encountered in the above case of 2-dimensional FFT. The Transpose
Algorithm intorduced above applies readily to this situation.

\begin{figure}[!htb]
  \centerline{\epsfxsize=.6\textwidth\epsfbox{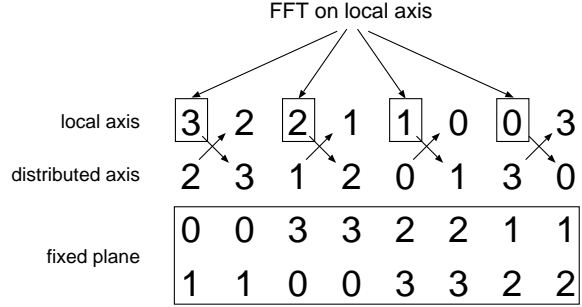}}
\caption[Layout]{
  Schematical view of 4-dimensional FFT on a 3-dimensional processor
  grid.
\label{FOUR}}
\end{figure}

The order of local axes 4-dimensional FFT is schematically illustrated
in \fig{FOUR}.
Initially, we perform the FFT along the third dimension, regarded as  local
at this time. Next, we transpose axis 2 with axis 3 and
perform the FFT along the second axis being local now. This
transposition is carried in parallel for all coordinates $x_{0}$ and
$x_{1}$.  It is obvious how the process can be continued. In the
second step, we fix dimension 0 and 3 and transpose axis 1 with 2.
Hence we can FFT axis 1, being local now. In the third step, we fix
dimension 2 and 3, transpose axis 0 and 1 and can perform FFT on axis
0 being local now.  Finally, we transpose axis 3 with 0 and re-arrive
in the original data layout.  \alg{QCDFFT} gives the pseudo-code
representation for the ensuing algorithm.
\begin{algo}{4-dimensional parallel FFT on 3-dimensional processor grid.}%
            {\label{QCDFFT}}{
for\= for\= for\= for\= for\= for\=\kill 
call 1-d-FFT(axis 3)\\ 
call transpose(axis 2 and 3)\\ 
call 1-d-FFT(axis 2) \\ 
call transpose(axis 1 and 2)\\ 
call 1-d-FFT(axis 1) \\ 
call transpose(axis 0 and 1)\\ 
call 1-d-FFT(axis 0) \\ 
call transpose(axis 3 and 0)}
\end{algo} 

\section{Summary}
\noindent
We have presented a novel scalable parallel algorithm for matrix
transposition and its application within FFT.  Universality of the method
is achieved by assignment of the data onto an abstract systolic ring that
can be readily mapped onto any parallel system. For mesh systems the use of
the hyper-systolic communication structure improves the communicational
complexity considerably.  In our scheme we can avoid computation of local
addresses.  Thus, in particular for SISAMD systems like APE100/Quadrics,
the implementation of efficient FFT became feasiblle.

We have implemented the  FFT scheme within a 2-dimensional simulation of
the Kraichnan model for passive advection.  Further, we suggested an
implementation of 4-dimensional FFT applied within lattice quantum
chromodynamics.  In the first application the FFT performs with an
efficiency of about 9 \% of the peak performace of the system, in a
general application we reach about 6 \%.  

On the European Teraflops system APEmille to come there is no restriction
as to local addressing.  Pre-skewing and Re-skewing will benefit mostly.
Coding will be straightforward and the computation significantly faster
(about an order of magnitude for a 512 node machine compared to on
APE100/Quadrics).  On APEmille, we expect to achieve a performance for FFT
of about 15 \% of the theoretical peak performance.

To set the numbers achieved on APE100 and expected on APEmille into
perspective we can compare with results from general purpose parallel
machines.  On the Cray T3E for an 8 node partition, the 3-dimensional FFT
on complex data performs with 10 \% of the peak performance using the
optimized scientific library\cite{VOGELSANG}.  On the Hitachi SR2201, a
$8k\times 8k$ complex data array is Fourier transformed with a speed of 30
\% of the machine's peak performance\cite{HAAS}.

Currently we are working on an efficient realization of 1-dimensional
FFT.

\paragraph{Acknowledgements.} 
The numerical tests in this work have been carried out on a 512-node
QH4 system at INFN, Pisa, Italy and the 256-node QH2 and 128-node QH1
at the University of Bielefeld, Germany.  We thank the staff of these
centers for their support.  We thank Dr. R. Vogelsang, Cray/ZAM
J\"ulich, Germany and Dr. P. Haas, RUS, Stuttgart, Germany for
benchmarking the Cray T3E and the Hitachi SR2201.  Th. L. is indebted
to Paolo Palazzari for many useful discussions and the construction of
the APE100/Quadrics cost function tables.  The work of Th. L. is
supported by the Deutsche Forschungsgemeinschaft DFG under grant No.\ 
Schi 257/5-1.  F. T.  thanks K.  S. and Th. L. for their friendly
hospitality at HLRZ/J\"ulich, where part of
this work was performed.\\[12pt]
\noindent{\bf Appendix}\\[12pt]
\noindent{\bf A: Skewing on SISAMD Computers}\\[12pt]
\noindent The first and last phases in the matrix transposition presented
above involve the representation of $\matrix{M}$ in skew order.  On systems
with indexed addressing facilities skewing is just a local address
computation. However, for SISAMD systems like APE100/Quadrics, skewing is
more difficult as its naive implementation spoils scalability.

\begin{algo}{Skewing for SISAMD systems.}%
            {\label{SKEWER}}{
for\= for\= for\= for\= for\= for\=\kill 
foreach cell $i=0:p-1$ $\in$ systolic array \\
\> for $q=0:l-1$ \\ 
\> \> $g=h\,{\mbox{\footnotesize mod}}_{2^q}$ \\
\> \> where $g_{q} > 2^{q}$\\
\> \> \> for $j=0:p-1$\\
\> \> \> \> $m_{j,i}=m_{[(j+2^{q}),i]}$ \\ 
\> \> \> end for\\
\> \> endwhere\\
\> end for \\ 
end foreach}
\end{algo} 

In order to solve this problem we have developed a systolic skewing
algorithm that avoids indexed addressing and shows logarithmic complexity.
Thus the method is well suited for SISAMD systems.

We start from the initial data configuration as illustrated in
\fig{LAYOUT_NORMAL}.  Let $p$ be a number $2^{l}$, $l\in {\bf N}$, and
let $h$ number the cells from $0$ to $p-1$ along the systolic
ring.  Taking $g_{q}=h\,{\mbox{\footnotesize mod}}_{2^q}$,
$q=0,\dots,l-1$, we can partition the cell array successively in
$\frac{p}{2}$ $\frac{p}{4}$, $\dots$, $2$ parts of equally numbered
sequences.  In each step we can impose a conditional statement that is
true {\tt{where}} $g_{q} > 2^{q}$. There a local memory shift
operation by a stride of $2^{q}$ is executed.  After l such steps, the
data array is represented in skew order.  The process is illustrated
as pseudo-code in \alg{SKEWER}.

\vspace*{12pt}
\noindent{\bf B. Skewing by Translation Invariance}\\[12pt]
\noindent 
An alternative to pre and re-skewing as presented above can be constructed
by usage of the translation invariance of the system.  The task is to
calculate a Fourier transformed function $\tilde{f}_{j}$ (we consider one
dimension) that is computed from the real space function $f_{i}$,
$i=0,\dots,N-1$ which is a vector $f_{i}=g_{(i+k){\mbox{mod}_N}}$,
translated by $k$ elements.  The Fourier transform
\begin{equation}
\tilde f_{j}=\sum_{i=0}^{N-1} f_{i}\,\exp{\frac{2\pi\imath\,ji}{N}}
\end{equation}
can be written in terms of $g$:
\begin{eqnarray}
\tilde f_{j}
&=&
\sum_{i=0}^{N-1} 
g_{(i+k){\mbox{mod}_N}}
\,\exp{\frac{2\pi\imath\,ji}{N}}\nonumber\\
&=&
\exp{-\frac{2\pi\imath\,jk}{N}}
\sum_{i=0}^{N-1} 
g_{(i+k){\mbox{mod}_N}}
\,\exp{\frac{2\pi\imath\,j(i+k)}{N}}\nonumber\\
&=&
\exp{-\frac{2\pi\imath\,jk}{N}}
\sum_{l=0}^{N-1} 
g_{l}
\,\exp{\frac{2\pi\imath\,jl}{N}}.
\end{eqnarray}
The translation in real space amounts to the application of a mode
dependent phase factor in Fourier space.  Similarly, a translated left hand
side is computed using corresponding phase factors.

A 2-dimensional FFT as given in \alg{FFT} can be performed on SISAMD
systems omitting 3 skewing steps. However, one re-skewing step
remains:
\begin{algo}{2-dimensional parallel FFT using translation invariance.}%
            {\label{TIFFT}}{
for\= for\= for\= for\= for\= for\=\kill 
call 1-d-FFT-pre-skew\\ 
call row-shift\\ 
call 1-d-re-skew-FFT-pre-skew \\ 
call row-shift\\
call re-skew}
\end{algo} 
We note that in the actual implementation of this method one has to
carefully address precision issues.\\[12pt]

\end{document}